\documentclass[11pt,dvips]{article}

\usepackage{epsfig,times} 
\usepackage{picinpar}
\usepackage{wrapfig}
\usepackage{floatflt}
\makeatletter
\renewcommand{\section}{\@startsection{section}{1}{0in}
        {0.4\baselineskip}{0.1\baselineskip}{\Large\bf}}
\renewcommand{\subsection}{\@startsection{subsection}{2}{0in}
        {0.25\baselineskip}{-\baselineskip}{\large\bf}}
\renewcommand{\subsubsection}{\@startsection{subsubsection}{3}{0in}
        {0.1\baselineskip}{-\baselineskip}{\normalsize\bf}}
\makeatother
\newcommand{\cat}{C{\small AT}}
\newcommand{\apj}{ApJ}
\newcommand{\app}{Astrop.~Phys.}
\newcommand{\nimpra}{Nucl.~Instrum.~Methods Phys.~Res., Sect.~A}
\newcommand{\jphysg}{J.~Phys.~G: Nucl.~Phys.}
\newcommand{\arcdeg}{\hbox{$^\circ$}}
\newcommand{\width}{\ensuremath{\mathit{width}}}

\newcommand{\size}{\ensuremath{\mathit{Q_{tot}}}}
\begin{document}

%
\makeatletter\newcommand{\ps@icrc}{
\renewcommand{\@oddhead}{\slshape{OG.2.2.03}\hfil}}
\makeatother\thispagestyle{icrc}
%
\markright{OG 2.2.03}
\begin{center}
%
{\LARGE \bf Large zenith-angle observations with the CAT Cherenkov imaging
  telescope}
\end{center}

\begin{center}
%
%
{\bf G.~Mohanty$^{1}$ for the \cat{} Collaboration}\\
{\it LPNHE, Ecole Polytechnique and IN2P3/CNRS, Palaiseau, France}
\end{center}

\begin{center}
{\large \bf Abstract\\}
\end{center}
\vspace{-0.5ex}
%
%
We present here results from large zenith-angle observations with the \cat{}
atmospheric Cherenkov imaging telescope, based on data taken on the Crab
Nebula and on the blazar Mk501 from 1996 onwards. From Monte Carlo
simulations, the threshold energy of the telescope is expected to vary from
about 250 GeV at zenith to about 2 TeV at a zenith angle of 60\arcdeg{}. The
lower source-fluxes due to the increased threshold energy are partly
compensated for by an increase in the effective collection area at large
zenith angles, thus allowing a significant extension of the dynamic range of
the \cat{} telescope, with a tolerable loss in sensitivity. We discuss
the implications for source detection and energy spectrum measurements.
%

\vspace{1ex}

%
%
\section{Introduction:}
\label{sec:intro}
The \cat{} atmospheric Cherenkov imaging telescope (Barrau et al., 1998)
has been operating from 1996 onwards at the site of a
former solar plant, Th\'{e}mis, in the French Pyr\'{e}n\'{e}es. It achieves a
low energy threshold of about 250 GeV at the zenith, in spite of a relatively
small mirror area of $17.7\hbox{ m}^2$, by discriminating against the
night-sky background with the rapid timing made possible by the
near-isochronous mirror, fast photomultipliers, and high-speed trigger and
readout electronics. It also uses a high-definition camera, with a central
region of 546 pixels of 0.12\arcdeg{} each, that allows the use of a powerful
$\chi^2$ image-analysis technique (LeBohec et al., 1998)
which very effectively rejects background shower
images while also offering excellent source location and energy resolution for
gamma-ray images. The \cat{} telescope has detected gamma-ray emission from
the Crab Nebula (Iacoucci et al., 1998) 
and from the
extra-galactic BL Lac objects, Mk501 (Djannati-Ata\"{\i} et al., 1999. Also,
see OG.2.1.08, these proceedings) and Mk421 (described in OG.2.1.09, these
proceedings). This paper discusses the use of the telescope for large
zenith-angle (here, $>$ 35\arcdeg{}, unless otherwise indicated) observations
on the Crab Nebula and on Mk501.

Large zenith-angle (LZA) observations of the Cherenkov light from air-showers
were originally proposed by Sommers \&{} Elbert~(1987)
as a means of increasing the effective detection area. The technique has since
been successfully applied to a variety of instruments~(e.g., Krennrich et al.,
1997, Krennrich et al., 1999, Tanimori et al., 1998)
Due to the greatly increased sensitivity at higher threshold energies, this
technique enables the study of the high-energy end of source spectra which is
interesting for various purposes, such as the modelling of the gamma-ray
emission mechanism at the source, measurement of the infra-red intergalactic
photon density, etc.

\section{The large zenith-angle technique}
\label{sec:lza-tech}
For large zenith-angle observations, the particle cascade in the air-shower
develops relatively farther away from the telescope, and furthermore, the
Cherenkov light from an air-shower is seen through a larger atmospheric depth.
This implies a higher detection energy threshold for a trigger based on the
photon density at ground level, which leads to a decrease in the observed flux
from observed TeV gamma-ray sources that exhibit spectra falling sharply with
energy. The lower source fluxes are compensated for by the increase in the
effective detector area resulting both from the effect of geometrical
projection onto the ground and from the shower being more highly developed at
the observatory level. Also, as the maximum of shower development is further
away from the telescope, the images at a given energy are smaller and fainter,
with a lower signal-to-noise ratio. This change in the image parameter
distributions need to be taken into account in the gamma-ray selection cuts.
The calibration of the detector at large zenith angles is also more
complicated as the data are more sensitive to fluctuating weather conditions.

\section{Simulations and data analysis}
\label{sec:sims}
Data were taken on the three sources above in the usual ON/OFF mode, where the
source is tracked for about 28 min., then the telescope is slewed back and is
used to observe a control region covering the same range of right-ascension
and declination. In general, more data are taken ON-source than OFF-source.
The data quality is constantly monitored for changing weather conditions and
electronic problems, and the conversion efficiency from Cherenkov photons at
ground level to photoelectrons is regularly measured. This enables us to
select data taken under stable observing conditions, with, typically, 75\%{}
of the data runs being deemed suitable for analysis. Here, the data are
analysed by two independent methods to guard against systematic effects at
large zenith angles. The first is the usual sophisticated $\chi^2$ analysis,
which automatically takes the effect of the zenith angle into account, with
the cuts being: $\hbox{P}\left(\chi^2\right) > 0.35$, $\alpha < 6\arcdeg{}$,
along with a cut on the total charge, $\size{} > 30\hbox{ pe}$, that rejects
low-light images. The second method is extended supercuts (Mohanty et al.,
1998),
which uses the usual Hillas parameters but allows them to scale with \size{}.
For the second method, simulated gamma-rays and background from data are used
to optimize the passbands for the various parameters at several discrete
zenith angles and these are then interpolated to all zenith angles. All
available data on the two sources up to the end of April 1999 has been
considered.

Gamma-ray showers have been simulated with the KASCADE shower simulation
program (Kertzman \&{} Sembroski, 1994),
modified locally by G.~Vacanti and E.~Pare. The response of the \cat{}
detector is simulated by other programs that include a comprehensive modelling
of the telescope optics and electronics. All simulations have been validated
on the strong gamma-ray signal observed from Mk501 in April 1997, when it was
at an extremely high-flux level.

\begin{figure}[bth] 
  \epsfig{
    bbllx=-75,bblly=280,bburx=567,bbury=510,
    file=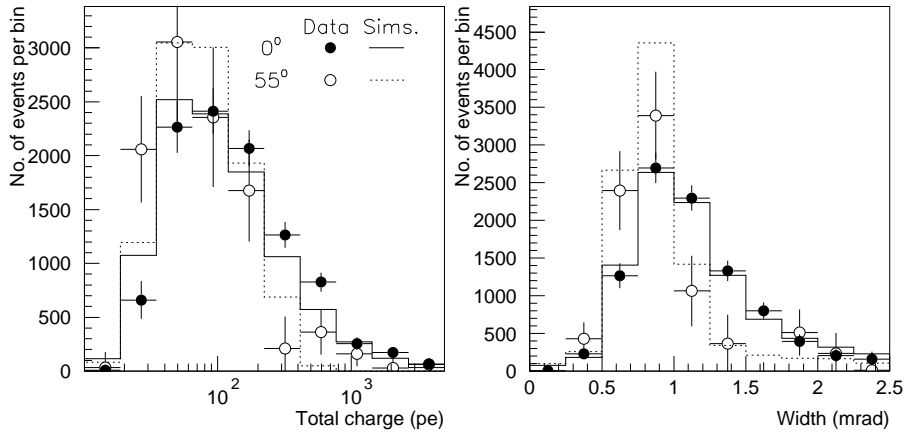, width=0.9 \linewidth}
  \caption{Comparison of parameter distributions from simulations and data at
    two different zenith angles.}
  \label{fig:par-dist}
\end{figure}

Simulated showers were generated with primary energies from 0.05 TeV to 10
TeV, drawn from a power-law distribution with a -2.25 differential index, and
scattered uniformly over a circle of 250 m radius.  Fig.~\ref{fig:par-dist}
compares the distributions of \size{} and the Hillas parameter, \width{}, from
the simulations, at 0\arcdeg{} and 55\arcdeg{}, to those from the excess in
data taken on Mk501 in a high state, centred on the corresponding zenith
angles.  Both simulations and data have been selected only by $\alpha <
6\arcdeg{}$, and all distributions have been renormalized to have the same
(arbitrary) total number of events. The agreement is seen to be good at both
zenith angles, though the statistics at 55\arcdeg{} are limited. The
distributions also illustrate the expected effect of fainter and smaller
images at larger zenith angles.

\begin{figure}[bht] 
  \epsfig{
    bbllx=-50,bblly=280,bburx=567,bbury=525,
    file=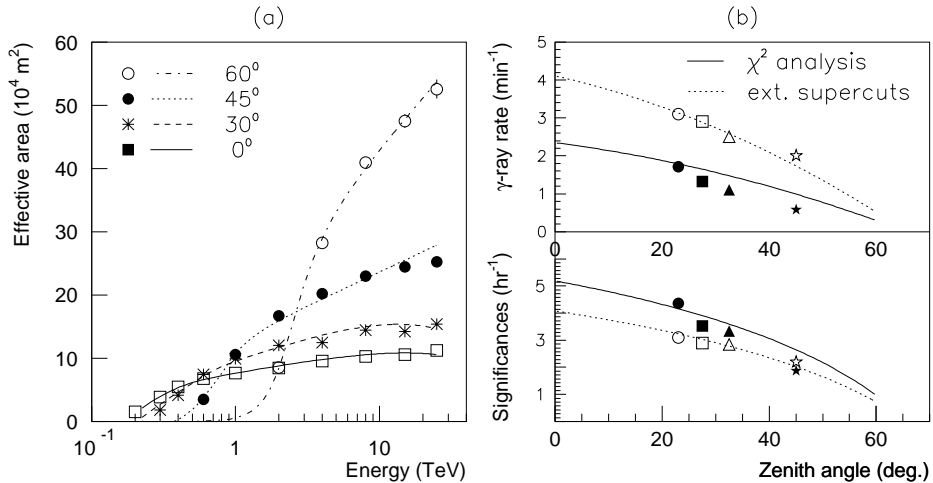, width=0.9 \linewidth}
  \caption{(a) Effective detector area at various zenith angles. (b)
    $\gamma$-ray rates (upper panel) and significances (lower panel) from a
    Crab-like source. The lines indicate calculated values while data from the
    Crab Nebula at various zenith angles are shown as points ($<25\arcdeg{}$:
    circles, $25\arcdeg{}$--$30\arcdeg{}$: squares,
    $30\arcdeg{}$--$40\arcdeg{}$: triangles, and $>40\arcdeg{}$: stars). The
    solid points are from the $\chi^2$ analysis,and the open ones from
    extended supercuts.}
  \label{fig:area-sig}
\end{figure}

Fig.~\ref{fig:area-sig}(a) shows the dramatic change in the effective detector
area, before the application of gamma-ray cuts, calculated from simulations at
several different zenith angles. The detector threshold energy is
conventionally defined as the peak of the expected telescope event rates for a
source with a power-law spectrum of differential index -2.7, being only weakly
dependent on the assumed index. The threshold energies thus defined, are, 0.24
TeV at 0\arcdeg{}, 0.35 TeV at 30\arcdeg{}, 0.70 TeV at 45\arcdeg{}, and 2.2
TeV at 60\arcdeg{}.  Fig.~\ref{fig:area-sig}(b) has two panels comparing the
gamma-ray rate, and the significance for a Crab-like source observed at
different zenith angles.  The lines indicate the expected values calculated
from the known background rate as a function of zenith angle, and from
gamma-ray simulations for a Crab-like source. The points are experimental
values for the excesses on the Crab obtained at various zenith angles. (As the
Crab culminates at about 20\arcdeg{} at Th\'{e}mis, the points are limited to
zenith angles above that.) The rates and significances are seen to be in good
agreement for both methods, except at very large zenith angles. Also, it
should be noted that the increase in detector area compensates to a large
extent for the increase in the energy threshold, the decrease in the gamma-ray
rate with zenith angle being much slower than would be otherwise expected.

\section{Results}
\label{sec:results}
\begin{figure}[thb] 
  \epsfig{
    bbllx=0,bblly=280,bburx=567,bbury=510,
    file=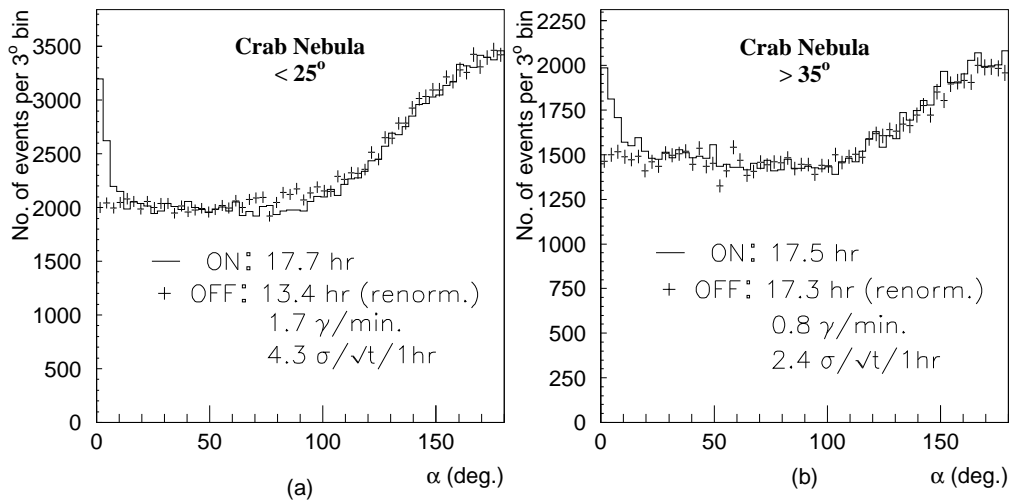, width=0.9 \linewidth}
  \caption{Excess from the Crab Nebula at (a) small, and (b) large zenith
    angles.}
  \label{fig:crab-alp}
\end{figure}

The simulations and data were shown to be in good agreement in the previous
section. Fig.~\ref{fig:crab-alp} presents the excess on the Crab Nebula, at
small ($<25\arcdeg{}$) and large ($>35\arcdeg{}$) zenith angles, in the
pointing angle, $\alpha$, after selection on $\hbox{P}\left(\chi^2\right)$ and
\size{}. The OFF data has been renormalized by the ratio between the
OFF-source and ON-source observation time. The excess at large zenith angles
is clearly evident. By further restricting the data to zenith angles above
40\arcdeg{}, one can examine the high-energy end of the flux from the Crab
Nebula. From just 9 hours of data above 40\arcdeg{}, we observe a 2$\sigma$
excess above a reconstructed energy of 10 TeV (this corresponds to a real
energy of 7 TeV at the 95\%{} confidence level, assuming a power-law spectrum
with a differential spectrum of -2.5). By comparison, one would need about ten
times as much observation time below 25\arcdeg{} to arrive at the same
significance at 10 TeV. Observations at large zenith angles also allow the
cross-calibration of spectra derived in different zenith angle bins.

\begin{figure}[bht] 
  \epsfig{
    bbllx=0,bblly=280,bburx=567,bbury=510,
    file=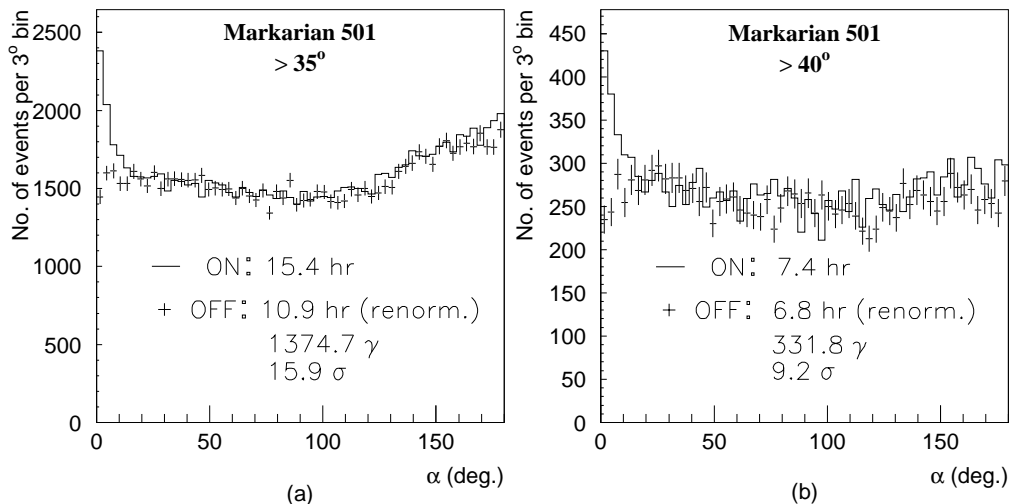, width=0.9 \linewidth}
  \caption{The large zenith-angle excess from Mk421: (a) above 35\arcdeg{},
    and, (b) above 40\arcdeg{}.}
  \label{fig:mk501-alp}
\end{figure}

Similarly, Fig.~\ref{fig:mk501-alp}(a) shows the LZA excess in $\alpha$ for
Mk501. The data on Mk501 is dominated by a period of high activity in 1997
which is seen also at very large zenith angles. This enables us to go down to
below 40\arcdeg{} for the flare period as shown in
Fig.~\ref{fig:mk501-alp}(b). Here also, the excess at large zenith angles is
easily visible. The spectrum of MK501 deviates considerably from a power-law
spectrum (see, e.g., Krennrich et al. 1999, Djannati-Ata\"{\i} et al., 1999).
This curvature is seen also in the LZA observations, and is presently under
investigation.
\section{Conclusions}
\label{sec:conclude}
We have demonstrated the ability of the \cat{} telescope to observe down to
large zenith angles, with the excess in the data agreeing fairly well with the
simulations. We are in the process of estimating energy spectra with the LZA
technique.
\vspace{1ex}
\begin{center}
{\Large\bf References}
\end{center}
%
Barrau, A., et al. 1998, \nimpra, 416, 278\hfill\\
Iacoucci, L., et al., 1998, in Proc.~$16^{\hbox{th}}$ European Cosmic Ray
Conference, 363\hfill\\
Kertzman, M., and Sembroski, G., 1994, \nimpra, 343, 629\hfill\\
Krennrich, F., et al., 1997, \apj, 482, 758\hfill\\
Krennrich, F., et al., 1999, \apj, 511, 149\hfill\\
Lebohec, S., et al., 1998, \nimpra, 416, 425\hfill\\
Mohanty, G., et al., 1998, \app, 9, 15\hfill\\        
Sommers. P., and Elbert, J.~W., 1987, \jphysg, 13, 553\hfill\\
Tanimori, T., et al., 1998, 492, L33\\

\end{document}